\documentclass[lettersize,journal]{IEEEtran}
\usepackage{amsmath,amsfonts}
\usepackage{algorithmic}
\usepackage{algorithm}
\usepackage{array}
\usepackage[caption=false,font=normalsize,labelfont=sf,textfont=sf]{subfig}
\usepackage{textcomp}
\usepackage{stfloats}
\usepackage{xcolor}
\usepackage{url}
\usepackage{verbatim}
\usepackage[pdftex]{graphicx}
\usepackage{booktabs}
\usepackage{cite}
\usepackage{multicol}
\usepackage{multirow}
\usepackage{tabularx}
\usepackage{threeparttable}
\begin{document}

\title{LPCM: Learning-based Predictive Coding for LiDAR Point Cloud Compression }
\author{Chang Sun, Hui Yuan,~\IEEEmembership{Senior Member,~IEEE,} Shiqi Jiang, Da Ai,~\IEEEmembership{Member,~IEEE,} Wei Zhang,~\IEEEmembership{Member,~IEEE,} Raouf Hamzaoui,~\IEEEmembership{Senior Member,~IEEE} 
\vspace{-1cm}
        % <-this % stops a space
\thanks{This work was supported in part by the National Natural Science Foundation of China under Grants 62222110, 62172259, and the High-end Foreign Experts Recruitment Plan of Chinese Ministry of Science and Technology under Grant G2023150003L, the Taishan Scholar Project of Shandong Province (tsqn202103001), the Natural Science Foundation of Shandong Province under Grant ZR2022ZD38, the Key Technology Research and Development Program of Shandong Province 2024CXGC010212.
\textit{(Corresponding author: Hui Yuan.)}}% <-this % stops a space
\thanks{Chang Sun, Hui Yuan and Shiqi Jiang are with the School of Control Science and Engineering, Shandong University, Jinan 250061, China, and also with the Key Laboratory of Machine Intelligence and System Control, Ministry of Education, Jinan 250061, China (e-mail: huiyuan@sdu.edu.cn).}
\thanks{Da Ai is with the School of Communication and Information Engineering, Xi'an University of Posts and Telecommunications, Xi’an, 710121, China.}
\thanks{Wei Zhang is with the School of Telecommunications Engineering, Xidian University, Xi’an, 710071, China}
\thanks{Raouf Hamzaoui is with the School of Engineering and Sustainable Development, De Montfort University, LE1 9BH Leicester, U.K.}
}
% The paper headers
\markboth{Journal of \LaTeX\ Class Files,~Vol.~14, No.~8, August~2021}%
{Shell \MakeLowercase{\textit{et al.}}: A Sample Article Using IEEEtran.cls for IEEE Journals}

% Remember, if you use this you must call \IEEEpubidadjcol in the second
% column for its text to clear the IEEEpubid mark.

\maketitle

\begin{abstract}
In recent years, LiDAR point clouds have been widely used in many applications. Since the data volume of LiDAR point clouds is very huge, efficient compression is necessary to reduce their storage and transmission costs. However, existing learning-based compression methods do not exploit the inherent angular resolution of LiDAR and ignore the significant differences in the correlation of geometry information at different bitrates. The predictive geometry coding method in the geometry-based point cloud compression (G-PCC) standard uses the inherent angular resolution to predict the azimuth angles. However, it only models a simple linear relationship between the azimuth angles of neighboring points. Moreover, it does not optimize the quantization parameters for residuals on each coordinate axis in the spherical coordinate system. To address these issues, we propose a learning-based predictive coding method (LPCM) with both high-bitrate and low-bitrate coding modes. LPCM converts point clouds into predictive trees using the spherical coordinate system. In high-bitrate coding mode, we use a lightweight Long-Short-Term Memory-based predictive (LSTM-P) module that captures long-term geometry correlations between different coordinates to efficiently predict and compress the elevation angles. In low-bitrate coding mode, where geometry correlation degrades, we introduce a variational radius compression (VRC) module to directly compress the point radii. Then, we analyze why the quantization of spherical coordinates differs from that of Cartesian coordinates and propose a differential evolution (DE)-based quantization parameter selection method, which improves rate-distortion performance without increasing coding time. Experimental results on the LiDAR benchmark \textit{SemanticKITTI} and the MPEG-specified \textit{Ford} datasets show that LPCM outperforms G-PCC and other learning-based methods. Specifically, LPCM achieves an average BD-Rate of -24.0\% compared to the G-PCC test model TMC13 v23.0 and -4.3\% compared to the current state-of-the-art learning-based method, SCP-EHEM.   
\end{abstract}
\vspace{-0.1cm}
\begin{IEEEkeywords}
Point cloud geometry compression, Predictive geometry coding, Rate-distortion optimization, Deep learning.
\end{IEEEkeywords}

\section{Introduction}
\IEEEPARstart{D}{ue} to its high accuracy and ease of recording 3D scenes and objects, LiDAR is widely used in automated driving \cite{1}, robotics navigation \cite{2}, geographic information systems \cite{3}, etc. Points recorded by LiDAR contain at least geometry information (e.g., Cartesian coordinates $x,y,z$) and sometimes attribute information (e.g., reflectance), and are collectively known as a point cloud. Since the data volume of LiDAR point clouds is very huge, efficient compression is necessary to reduce their storage and transmission cost. However, point clouds are sparse and irregularly distributed in 3D space, making their efficient compression more challenging compared to structured data like images or videos.

The Moving Picture Experts Group (MPEG) recently released the geometry-based point cloud compression (G-PCC) \cite{4} standard and is actively researching new point cloud compression techniques. G-PCC uses either octree coding \cite{5} or predictive geometry coding \cite{7}\cite{8} to encode LiDAR point cloud geometry information. As the predictive geometry coding method is designed based on the LiDAR acquisition principle, it outperforms the octree coding method. LiDAR captures point clouds using multiple lasers with fixed elevation angles and an inherent angular resolution. This acquisition principle results in stronger correlations between points in the spherical coordinate system. To leverage this acquisition principle, the angular mode \cite{9}\cite{10} in the predictive geometry coding method converts geometry information into the spherical coordinate system and groups points based on their laser IDs. Within each group, points are connected in the order of their azimuth angles to construct a predictive tree. Then, starting from the root point, the geometry information is sequentially predicted using the encoded neighboring points, and the residuals are entropy encoded.

However, there are weaknesses in the predictive geometry coding method. First, this method predicts the geometry information of the current point by modeling a simple linear relationship between neighboring points in the predictive tree, neglecting the long-term geometry correlation among different coordinates in the spherical coordinate system. Second, due to the degradation of geometry correlation at low bitrates, it does not outperform the octree coding method at low bitrates. Finally, the distortion caused by the quantization of residuals on each axis in spherical coordinates contributes differently to the reconstruction error. However, the method does not address the rate-distortion (RD) optimization problem for the quantization parameters (QPs) of the spherical coordinate system, and thus it cannot achieve optimal RD performance. 

Recently, deep learning-based coding methods have shown excellent RD performance \cite{11}. For lossy coding, autoencoder-based methods \cite{12,13,14,15} are widely used. These methods use a hyper-prior-based entropy model to entropy encode the latent representation extracted from geometry information, which is efficient for dense point clouds. For lossless coding, voxel-based methods and octree-based methods are widely used. Voxel-based methods \cite{16}\cite{17} compress the point cloud by using an entropy model to predict the occupancy probability of each voxel, which is also efficient for voxel-based dense point clouds. Octree-based methods \cite{18,19,20,21,22,23,24} use learning-based entropy models to estimate the occupancy probability of octree nodes. These methods are suitable for both dense point clouds and sparse point clouds. However, they do not account for the acquisition principle of LiDAR, resulting in lower compression efficiency compared to the predictive geometry coding method of G-PCC. Luo et al. \cite{25} proposed SCP-EHEM, which represents a point cloud using a spherical-coordinate-based octree and encodes it with EHEM \cite{23}. However, SCP-EHEM only learns the distribution of LiDAR point clouds in the spherical coordinate system to build efficient entropy models, neglecting LiDAR's inherent angular resolution, which limits its RD performance.

Motivated by the above analysis, we propose a learning-based predictive coding method (LPCM) for the geometry of LiDAR point clouds. LPCM first transforms point clouds into predictive trees in a spherical coordinate system. At high bitrates, following the order of the points in the predictive tree, LPCM predicts the geometry information of each point using a Long-Short-Term Memory-based predictive (LSTM-P) module. The quantized residuals are then entropy encoded. At low bitrates, as the geometry correlation decreases, LPCM directly encodes the radius using a variational radius compression (VRC) module. Compared to  the predictive geometry coding of G-PCC, LPCM can exploit the long-term correlation of geometry information between different coordinates. Compared to other learning-based methods, LPCM takes advantage of the acquisition principle and inherent angular resolution of LiDAR to eliminate geometry redundancy. The contributions of our work can be summarized as follows.

\begin{enumerate}
\item {We propose LPCM, a learning-based predictive coding method for the geometry of LiDAR point clouds. LPCM includes high-bitrate and low-bitrate coding modes to achieve efficient compression across various bitrates.}
\item {At high bitrates, we propose an LSTM-P module to predict the geometry of points by capturing the long-term geometry correlation between different coordinates. At low bitrates, we achieve more efficient compression by directly encoding the radius using the proposed VRC module.}
\item {To further improve RD performance, we analyze the reconstruction error caused by quantization on each axis in the spherical coordinate system and propose a differential evolution (DE)-based QP selection method to determine the QPs for residuals.}
\item {Experimental results on the LiDAR benchmark \textit{SemanticKITTI} and the MPEG-specified \textit{Ford} datasets show that LPCM achieves a BD-Rate of -24.0\% compared to the G-PCC test model TMC13 v23.0 and -4.3\% compared to the current state-of-the-art method, SCP-EHEM, across a wide range of bitrates (from 0.9 bits per input point to 10.5 bits per input point). Moreover, the reconstructed point cloud yields the best performance in vehicle detection.}
\end{enumerate}

The remainder of this paper is organized as follows. Section 2 provides a brief overview of existing geometry coding methods. Section 3 describes the proposed method in detail. Section 4 presents and analyzes the experimental results. Finally, Section 5 concludes the paper and suggests possible future work.

\section{Related work}
In this section, we review three relevant areas: geometry compression methods in the G-PCC coding standard, deep learning-based geometry compression methods, and RD-optimized QP selection for geometry compression.

\subsection{Geometry Compression in G-PCC}
MPEG proposed the first edition of G-PCC \cite{4} in 2021 and is currently developing the second edition\cite{6}. G-PCC includes three point cloud geometry encoding methods: octree coding method \cite{5}, trisoup coding method \cite{29}, and predictive geometry coding method \cite{7}\cite{8}. Trisoup relies on a dense and continuous surface to form triangles, making it less effective for LiDAR point clouds. The octree coding method uses an octree structure to represent a voxelized point cloud and encodes the occupancy information of each node by an adaptive arithmetic coder. This method can efficiently compress dense point clouds. However, its performance declines when compressing sparse LiDAR point clouds. The predictive geometry coding method connects discrete points into a predictive tree, sequentially predicts the geometry coordinates of each point, and encodes the coordinate residuals. The angular mode \cite{9} in the predictive geometry coding method constructs a predictive tree based on the laser IDs and azimuth angles of the points. This method outperforms the octree coding method for LiDAR point cloud compression by leveraging the inherent angular resolution of LiDAR. Yu et al. \cite{30} observed irregular point distribution in the angular mode and proposed a predictive relationship between coordinates. However, this method ignores LiDAR acquisition errors and environmental influences. Angle adaptive quantization \cite{10}, which adaptively quantizes the azimuth angles based on the radius, further improves coding efficiency. However, this quantization method only relates the radii and the QP of the azimuth angles, without considering the relationship between the QP of the radius and that of the angle. Because each axis in spherical coordinates contributes differently to the reconstruction error, optimal RD performance cannot be achieved.

\subsection{Learning-based Geometry Compression}
For lossy coding, Huang et al. \cite{31} proposed an autoencoder-based method for point cloud compression. Specifically, the latent representation generated by the encoder is compressed using entropy coding. The decoder then reconstructs the point cloud from this latent representation. Quach et al. \cite{32} proposed a joint loss function of both rate and distortion by using a trade-off parameter. The works in \cite{12,13,14,15} use entropy models to predict the probability distribution of latent representations extracted by an encoder, thereby enhancing the compression efficiency of entropy coding. Wu et al. \cite{33} leveraged human geometric priors to remove geometry redundancy in point clouds of human models, significantly improving compression efficiency. For inter-frame coding, Akhtar et al. \cite{37} proposed a predictive network that uses the previous frame to predict the latent representation of the current frame and encodes the residuals. However, due to the sparsity and irregularity of LiDAR point clouds, these methods are not suitable for compressing them. 

For lossless coding, point clouds are typically voxelized first. Then, the current voxel is entropy encoded based on the occupancy probability predicted from the encoded voxels (0: empty voxel, 1: occupied voxel). Both VoxelDNN \cite{16} and CNet\cite{17} use an entropy model to predict the occupancy probability of the current voxel, compressing the point cloud voxel by voxel. However, this method requires point by point decoding, resulting in high complexity. To reduce the coding complexity, MS-VoxelDNN \cite{35} organizes voxels within a multiscale architecture and models occupancy at each scale in a coarse-to-fine order. SparsePCGC \cite{36} compresses scale-wise occupied voxels by exploiting cross-scale and same-scale correlations flexibly. For inter-frame coding, Wang et al. \cite{38} use lower-scale priors from both the current frame and encoded frames to enhance the conditional probability estimation for efficient entropy coding. Similar to autoencoder-based methods, these voxel-based methods efficiently compress dense voxelized point clouds, but are not effective for compressing LiDAR point clouds. 

As an octree is generally used to represent voxelized point clouds, researchers have proposed predicting the occupancy probability (with a total of 255 possibilities) of the current node based on already encoded octree nodes, followed by entropy coding. Huang et al. \cite{18} proposed a context model called OctSqueeze that uses ancestor nodes as context to predict the probability distribution of node occupancy. MuSCLE \cite{19} introduces an inter-frame context that incorporates ancestor nodes and neighboring nodes from the previously encoded frame. OctAttention \cite{21} introduces a large-scale context and uses an attention mechanism to emphasize the correlated nodes in the context. However, this method requires decoding the octree node by node, resulting in high decoding complexity. Song et al. \cite{23} introduced a grouped context structure to avoid node by node decoding while preserving the compression performance. In our previous work \cite{24}\cite{34}, we found that the cross-entropy loss function fails to accurately measure the difference between the one-hot encoding and the predicted probability distribution. To address this, we proposed an attention-based child node number prediction module to assist the context model in predicting the probability distribution for efficient entropy coding. However, similar to the octree coding method in G-PCC, the performance of these methods declines when compressing sparse LiDAR point clouds. Luo et al. \cite{25} proposed SCP-EHEM, which represents a point cloud using a spherical-coordinate-based octree and then encodes the spherical octree using EHEM \cite{23}. While this method considers the LiDAR acquisition principle, it only learns the distribution of LiDAR point clouds in the spherical coordinate system, and does not account for the inherent angular resolution of LiDAR.

Several methods were specifically designed for LiDAR point clouds. Song et al. \cite{39} proposed a layer-wise geometry aggregation (LGA) framework for lossless geometry compression. Fan et al. \cite{40} used progressive downsampling to partition point clouds and proposed a fully factorized entropy model that exploits spatial correlations at each level to compress the latent variables effectively. Zhao et al. \cite{41} used semantic prior representation (SPR) for point clouds and proposed a lossy coding algorithm with variable precision to generate the final bitstream. However, these methods did not exploit the acquisition principle and inherent angular resolution of LiDAR, which limits the RD performance.

\subsection{RD-Optimized Quantization Parameter Selection}
Existing RD optimization typically constructs a model that represents the relationship between encoding parameters (i.e., QP, octree depth) and RD performance to determine the encoding parameters. Liu et al. \cite{42} used the relationship between octree levels, bitrate, and attribute distortion to formulate the RD optimization problem as a constrained convex optimization problem and apply an interior point method to solve it. Li et al. \cite{43} modeled the relationship between geometry bitrate and geometry QP and proposed a frame-level bit allocation algorithm. The above methods use the same QP to quantize different coordinate axes in the Cartesian coordinate system and then analyze their RD performance. However, the distortion on each coordinate axis in the spherical coordinate system affects the reconstruction quality differently. Applying the same QP cannot yield optimal RD performance. Hou et al. \cite{44} modeled the relationship between QPs and bitrates in the spherical coordinate system. However, this method determines QPs based on the visually observed pattern that a larger difference between the QP of the radius and that of the angular leads to better RD performance, without accurately modeling the relationship between QPs and distortion.  

In summary, existing learning-based methods do not exploit the acquisition principle and inherent angular resolution of LiDAR and ignore the significant
differences in correlation of geometry information at different bitrates. Predictive geometry coding in G-PCC leverages the inherent angular resolution, outperforming most learning-based methods in LiDAR point cloud compression. However, it only predicts the geometry information of the current point by modeling a simple linear relationship between neighboring points in the predictive tree, without exploiting long-term geometry correlation between different coordinates. Moreover, due to the lack of accurate modeling of the relationship between the QP of the radius and that of the azimuth angle, current RD optimization methods \cite{10}\cite{44} are unable to obtain optimal QPs.

\section{Proposed Method}

The architecture of the proposed LPCM is shown in Fig. 1. First, we convert the input point cloud into predictive trees in the spherical coordinate system. Next, we design high-bitrate and low-bitrate coding modes, respectively, by observing that the correlation between different coordinates decreases significantly at low bitrates. In the high-bitrate coding mode, the azimuth angles and radii are encoded using a differential coding (DC)-based encoder. Then, we use the proposed LSTM-P module, which captures long-term geometry correlations between different coordinates, to compress the elevation angles. In the low-bitrate coding mode, the radius is encoded using the proposed VRC module, while azimuth and elevation angles are encoded using a DC-based encoder. We also analyze how the distortion on each axis contributes differently to the reconstruction error and determine the QPs for azimuth angle, elevation angle, and radius using DE to improve RD performance.  

\begin{figure*}[!t]\centering
  \includegraphics[width=18cm]{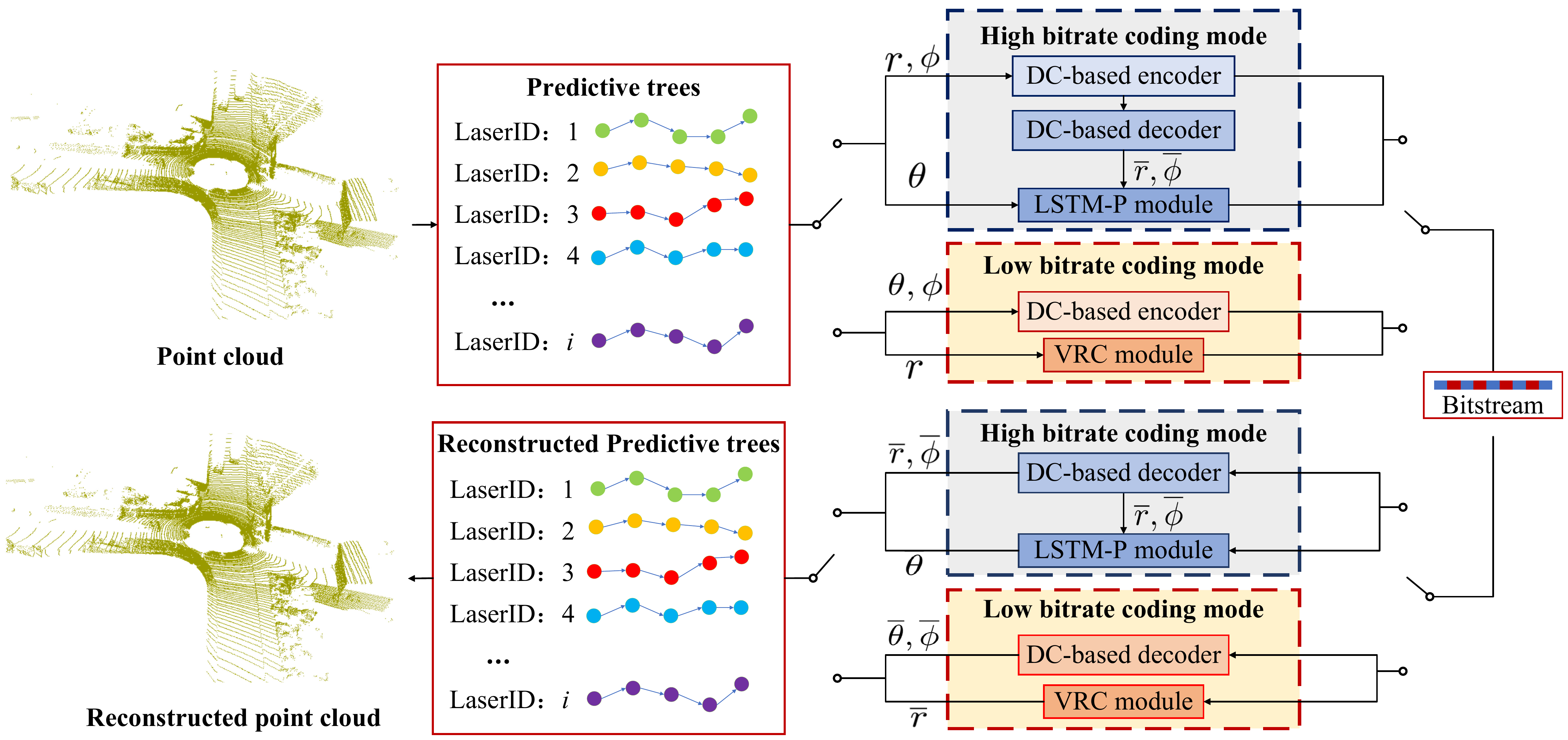}
  \caption{Overall architecture of the proposed LPCM. LPCM first converts the input point cloud into predictive trees. At high bitrates, the azimuth angles $\phi$ and the radii $r$ are compressed using a DC-based encoder. Then, the proposed LSTM-P module is used to compress the elevation angles $\theta$. At low bitrates, the elevation angles $\theta$ and the azimuth angle $\phi$ are compressed using a DC-based encoder, while the radii $r$ are compressed using the proposed VRC module.}
  \label{fig4}
\vspace{-0.3cm} 
\end{figure*}

\subsection{Predictive Tree Construction}
Due to variations in the annotation of calibration parameters of LiDAR and the arrangement of points in LiDAR point cloud files, we use two predictive tree construction methods. 

\begin{figure}[!t]
\centering
  \includegraphics[width=8cm]{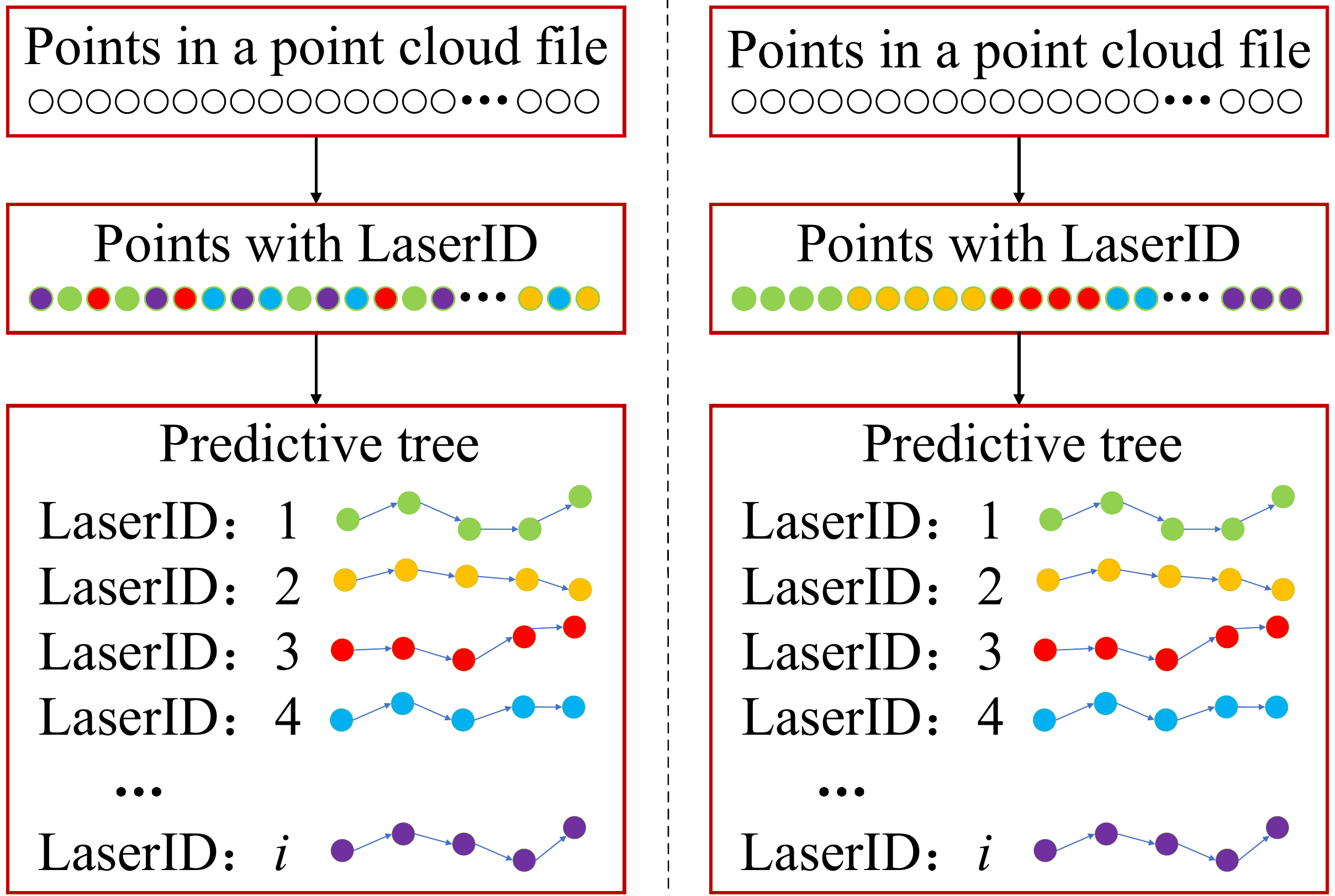}
  \vspace{-0.3cm}
  \caption{Construction of the predictive tree. The left side represents the calibration parameter-based method, while the right side represents the threshold segmentation method. White circles indicate points with an undetermined laser ID, whereas colors other than white correspond to different laser IDs.}
\label{fig100}
\vspace{-0.5cm} 
\end{figure}

As shown in the left part of Fig. 2, for point cloud files that contain the elevation angles and vertical heights of LiDAR lasers such as the \textit{\textit{Ford}} dataset \cite{49}, we use the G-PCC calibration parameters-based method to construct the predictive tree. We first convert the coordinates of the points from the Cartesian coordinate system to the spherical coordinate system
\begin{equation}
r = \sqrt{x^2 + y^2},
\end{equation}
\begin{equation}
\phi = \arctan2 (\frac{{y}}{x}),
\end{equation}
\begin{equation}
i = \arg\min_{j=1,\ldots,N} \left\{ z + \zeta(j) - r \cdot \tan\left(\theta(j)\right) \right\},
\end{equation}
where $(x,y,z)$ are the Cartesian coordinates of a point, $r$ is the radius, $\phi$ is the azimuth angle, $j$ is a candidate laser ID, $i$ is the actual laser ID, $N$ is the number of lasers, $\zeta(j)$ is the height of the $j$-th laser, and $\theta(j)$ is the elevation angle of the $j$-th laser. Both $\zeta(j)$ and $\theta(j)$ are provided in the \textit{Ford} dataset. Then, the elevation angle of the point is obtained as
\begin{equation}
\theta = \arctan2 (\frac{z + \zeta(i)}{r}).
\end{equation}
Finally, the points with the same laser ID are connected in the order of their azimuth angles to construct a predictive tree.

As shown in the right part of Fig. 2, for point cloud files without elevation angles and LiDAR laser vertical heights (such as the \textit{SemanticKITTI} dataset), the points must already be organized according to the order of laser IDs and their acquisition order. In this case, we construct the predictive trees using a threshold segmentation method. We first obtain the coordinates of the points in the spherical coordinate system. Next, we determine $\phi$ and $r$ using (1) and (2). $\theta$ can then be calculated as
\begin{equation}
\theta = \arctan2(\frac{z}{r}),
\end{equation}

As the azimuth angle cyclically varies from 1 $^\circ$ to 360 $^\circ$ and points originating from the same laser are arranged adjacently in ascending order of azimuth angles, identifying the boundary between points from two different laser scanners is sufficient to determine the laser ID for each point. When the difference in azimuth angle between point $p_n$ and point $p_{n-1}$ is greater than a preset threshold $t$, grouping is done as follows
\begin{equation}
i_n =
\begin{cases} 
j+1 & if\text{  } |\phi_n - \phi_{n-1}| \geq t \\
j & otherwise ,
\end{cases}
\end{equation}
where $i_n$ is the laser ID of point $p_n$, and $j$ is the laser ID of point $p_{n-1}$. Then, we can construct predictive trees based on the laser IDs and azimuth angles.

\subsection{High-bitrate Coding Mode}

After constructing the predictive tree, we first encode the root point and sequentially predict the coordinates of the following points. Then, the quantized residuals are compressed using entropy coding.

\textbf{Compression of azimuth angles.} The LiDAR acquisition principle ensures that the difference in azimuth angles between adjacent points in the predictive tree is represented by a multiple of a unit azimuth angle, i.e.,
\begin{equation}
\phi_n = \phi_{\text{unit}} \times s_n + \delta_n,
\end{equation}
where $s_n$ is an integer slope, $\delta_n$ is a bias, and $\phi_{\text{unit}}$ is a unit azimuth angle that can be expressed as
\begin{equation}
\phi_{\text{unit}} = \frac{\phi_{\text{ar}}}{q_{\phi}}.
\end{equation}
In (8), $\phi_{\text{ar}}$ is the inherent angular resolution, and $q_{\phi}$ is a QP. The bias $\delta_n$ is quantized as
\begin{equation}
\tilde{\delta}_n = \left\lfloor \delta_n \times q_\delta \right\rceil,
\end{equation}
where $q_\delta$ is a QP. The integers $\{s_1,s_2,...,s_n\}$ are encoded using DC followed by a context-adaptive binary arithmetic coding (CABAC). Similarly, the quantized biases $\{\tilde{\delta}_1, \tilde{\delta}_2,...,\tilde{\delta}_n\}$ are encoded using CABAC. At the decoder, the reconstructed azimuth angle $\overline{\phi}_{n}$ is obtained by

\begin{equation}
\overline{\phi}_{n} = {\phi_{\text{unit}}} \times {s_n} + \frac {\tilde{\delta}_n}{q_\delta}.
\end{equation}

\textbf{Radius compression.} The radius $r_n$ of point $p_n$ is predicted by the reconstructed radius $\overline{r}_{n-1}$ of point $p_{n-1}$, getting
\begin{equation}
\hat{r}_n = \overline{r}_{n-1}.
\end{equation}
Then the residual $res_{r,n}$ and quantized residual $\widetilde{res}_{r,n}$ is expressed as
\begin{equation}
res_{r,n} = r_n - \hat{r}_n,
\end{equation}
\begin{equation}
\widetilde{res}_{r,n} = \left\lfloor{res_{r,n}}\times{q_r} \right\rceil,
\end{equation}
where $q_r$ is the QP of the radius. Finally, $r_1$ and $\{\widetilde{res}_{r,2},...,\widetilde{res}_{r,n}\}$ are encoded using CABAC. At the decoder, the reconstructed radius $\overline{r}_{n}$ is obtained by
\begin{equation}
\overline{r}_{n} = \hat{r}_n + \frac{\widetilde{res}_{r,n}}{q_r}.
\end{equation}
\begin{figure}[!t]
\centering
  \includegraphics[width=9.5cm]{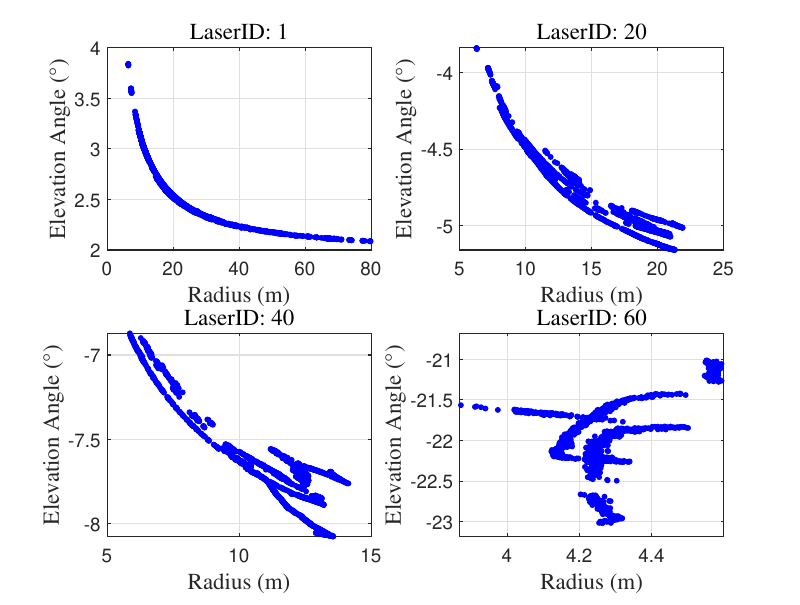}
  \vspace{-0.7cm} 
  \caption{Visualization of the elevation angle and radius of points from laser ID 1, 20, 40, and 60. The correlation between the elevation angle and the radius of points from different lasers varies significantly.}
\label{fig2}
\vspace{-0.1cm} 
\end{figure}

\textbf{Compression of elevation angles.} Since the elevation angle of a laser is fixed, points originating from this laser ideally share identical elevation angles. However, due to internal LiDAR calibration, acquisition errors, and environmental influences, the elevation angles of points from a laser often exhibit variation. We observed a significant correlation between radius and elevation angle, as shown in Fig. 3. The correlation varies across lasers with different elevation angles. Specifically, when the elevation angle of the laser is relatively large, there is a clear correlation between the elevation angle and the radius. As the elevation angle of the laser decreases, this correlation weakens. When the elevation angle becomes very small, we cannot observe any noticeable correlation. As a result, a simple function such as the one proposed in \cite{30} cannot accurately model the correlation between elevation angle and radius. 
\begin{figure}[!t]
\centering
  \includegraphics[width=9cm]{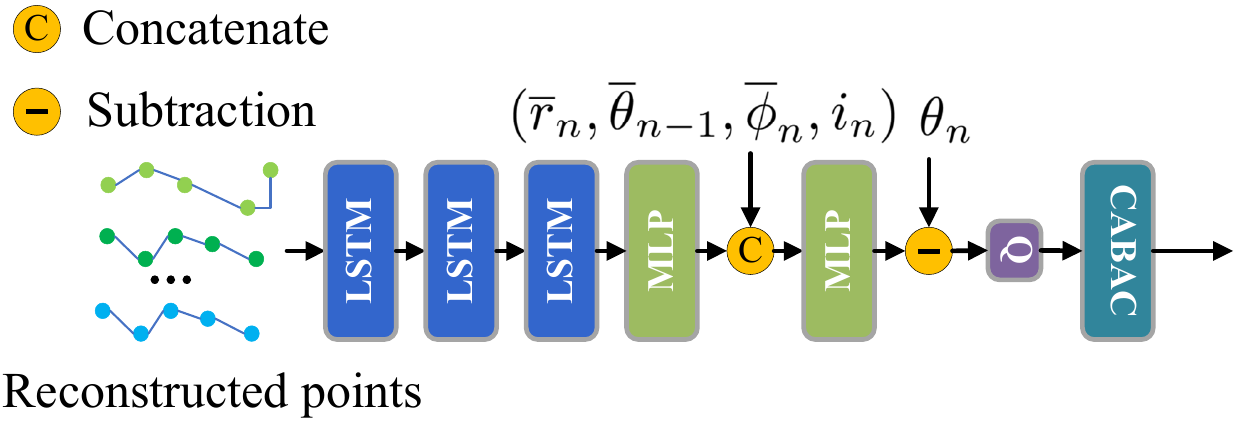}
  \vspace{-0.7cm} 
  \caption{Architecture of the proposed LSTM-P module. It consists of three LSTM layers and two MLPs. The input of the LSTM-P module consists of the reconstructed coordinates $(\overline{r}, \overline{\theta}, \overline{\phi})$ of a preset number of neighboring points in the predictive tree and the corresponding laser IDs $i$. The ouput of the first MLP is concatenated with the reconstructed elevation angle $\overline{\theta}_{n-1}$ of point $p_{n-1}$, the reconstructed azimuth angle $ \overline{\phi}_n$, radius $\overline{r}_n$, and laser ID $i_n$. Then, the second MLP outputs the predicted elevation angle $\hat{\theta}_{n}$. Finally, the residual between $\theta_{n}$ and $\hat{\theta}_{n}$ is quantized and encoded by CABAC.  }
\label{fig3}
\vspace{-1cm} 
\end{figure}

\begin{figure*}[!t]\centering
  \includegraphics[width=17cm]{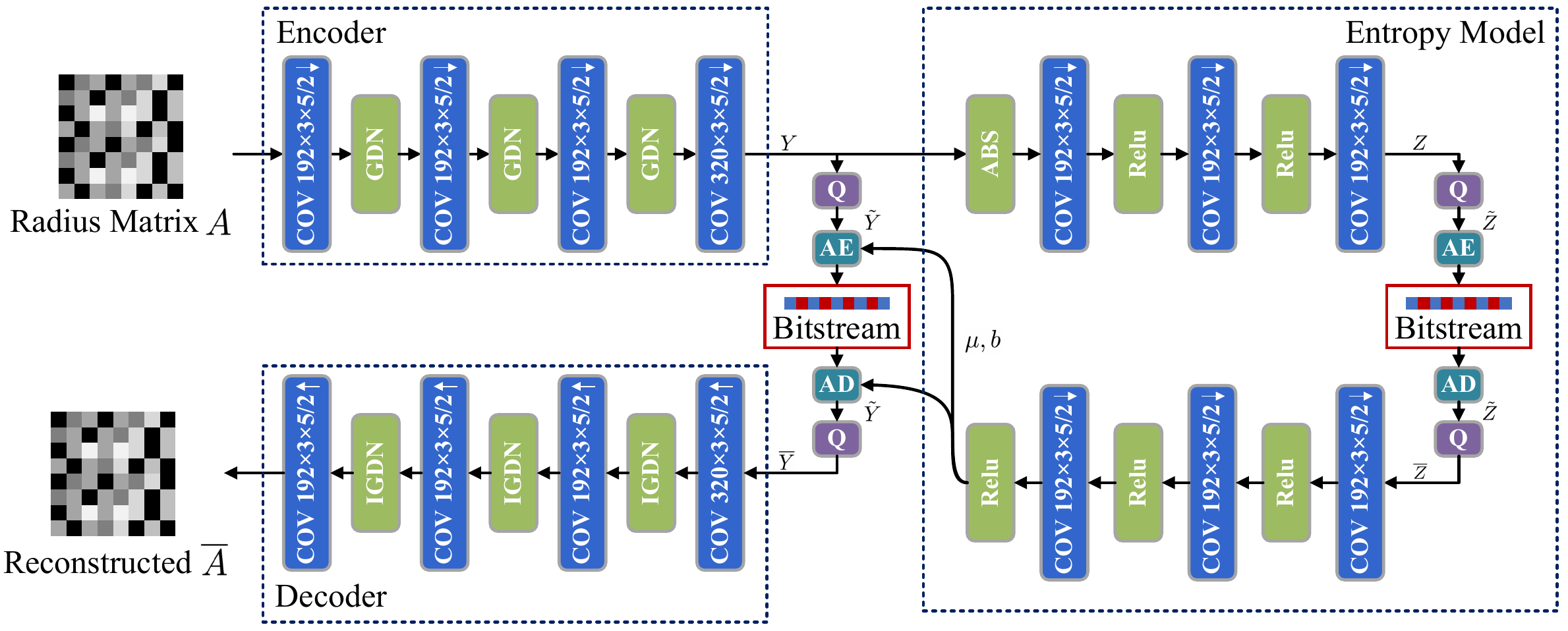}
  \vspace{-0.3cm}
  \caption{Overall architecture of the proposed VRC module. The left side is an autoencoder, while the right side is an entropy model. Q denotes quantization, and AE and AD denote an arithmetic encoder and arithmetic decoder, respectively. Convolution parameters are expressed as number of filters × height of kernel × width of kernel / upsampling or downsampling stride, where ↑ represents upsampling and ↓ represents downsampling. The encoder maps the matrices $\mathbf{A}$ to a latent representation $\mathbf{Y}$.  Then, the decoder maps the reconstructed latent representation $\mathbf{\overline{Y}}$ back to the reconstructed matrices $\mathbf{\overline{A}}$. The entropy model predicts the distribution of $\mathbf{\tilde{Y}}$ based on the hyper-prior information $\mathbf{Z}$ to enhance the efficiency of arithmetic encoding.}
  \label{fig4}
  \vspace{-0.5cm}
\end{figure*}

To efficiently compress elevation angles, we propose the LSTM-P module to capture the correlation between radius and elevation angle. As shown in Fig. 4, the proposed LSTM-P module consists of three LSTM layers and two multilayer perceptrons (MLPs). The reconstructed coordinates $(\overline{r}, \overline{\theta}, \overline{\phi})$ of a preset number of neighboring points in the predictive tree and the corresponding laser IDs $i$ are input to the LSTM. The reconstructed coordinates of previous points provide long-term geometry correlation between different coordinates, while the laser IDs contain the information of the relationship between radius and elevation angles as shown in Fig. 3. The output of the first MLP is concatenated with $(\overline{r}_n , \overline{\theta}_{n-1}, \overline{\phi}_n, i_n)$. Here, $\overline{\theta}_{n-1}$ is the reconstructed elevation angle of the point $p_{n-1}$, while $\overline{r}_n$ and $\overline{\phi}_n$ denote the reconstructed radius and azimuth angle of the current point $p_{n}$, respectively. Since $\overline{r}_n$ and $\overline{\phi}_n$ are encoded before the elevation angle, they are available for encoding $\theta_{n}$. $i_n$ is the laser ID of the current point, which can be obtained from the previous points. Accordingly, the second MLP outputs the predicted elevation angle $\hat{\theta}_{n}$. The residual $res_{\theta,n}$ and quantized residual $\widetilde{res}_{\theta,n}$ can then be expressed as
\begin{equation}
res_{\theta,n} = \theta_n - \hat{\theta}_n,
\end{equation}
\begin{equation}
\widetilde{res}_{\theta,n} = \left\lfloor {res_{\theta,n}}\times{q_\theta} \right\rceil,
\end{equation}
where $q_\theta$ is the QP of the elevation angle. Finally, ${\theta}_{1}$ and $\{\widetilde{res}_{\theta,2},\widetilde{res}_{\theta,3},...,\widetilde{res}_{\theta,n}\}$ are encoded using CABAC. At the decoder, the reconstructed elevation angle $\overline{\theta}_{n}$ is expressed as
\begin{equation}
\overline{\theta}_{n} = \hat{\theta}_n + \frac{\widetilde{res}_{\theta,n}}{ q_\theta},
\end{equation}

The LSTM-P module is trained with the mean squared error (MSE) loss between the actual elevation angle $\theta_{n}$ and the predicted elevation angle $\hat{\theta}_{n}$:
\begin{equation}
\mathcal{L}_{mse}=\frac1{N}\sum_{n=1}^{N}{\left\|\theta_{n}- \hat{\theta}_{n}\right\|_2^2}, 
\end{equation}
where $N$ is the total number of points.

\subsection{Low-bitrate Coding Mode}
At low bitrates, the significant distortion of the reconstructed coordinates prevents the LSTM-P module from accurately predicting elevation angles. Moreover, due to the wide range of radius values and the unclear variation patterns, the radius cannot be accurately predicted, resulting in a high bitrate for encoding the residuals. Therefore, at low bitrates, we use the proposed VRC module to reduce the bitrate. 

First, the azimuth angles are compressed using the same method as in the high-bitrate coding mode. Then, the predicted elevation angle $\hat{\theta}_{n}$ can be expressed as
\begin{equation}
\hat{\theta}_{n} = \overline{\theta}_{n-1},
\end{equation}
and the quantized residuals $\widetilde{res}_{\theta,n}$ are obtained using (15), (16), (17) and then entropy encoded using CABAC.

The radius is directly compressed using the proposed VRC module. As shown in Fig. 5, the VRC module consists of an autencoder and a hyper-prior entropy model. The autoencoder is composed of stacked convolution neural networks (CNNs) \cite{46} and generalized divisive normalization layers (GDNs) \cite{47}. The hyper-prior entropy model is composed of stacked CNNs and ReLU activation functions. The radii are first arranged according to the laser IDs and the order of the predictive tree, and then reshaped into several 256×256 matrices, detnoted as $\mathbf{A}$. As $\mathbf{A}$ is constructed in a row-first order, the correlation within the same row of $\mathbf{A}$ is higher. Therefore, all CNNs in the VRC module use rectangular convolutions with kernel sizes of $3\times5$ or $1\times3$. First, the encoder maps the matrices $\mathbf{A}$ to a latent representation $\mathbf{Y}$, which is then quantized to $\mathbf{\tilde{Y}}$. Then, we use a hyper-prior entropy model to estimate each element $\tilde{y}_m$ in $\mathbf{\tilde{Y}}$ using a Laplace distribution
\begin{equation}
p(\tilde{y}_m | \mathbf{\tilde{Z}}) = \frac{1}{2b_m} \exp\left(-\frac{|\tilde{y}_m - \mu_m|}{b_m}\right),
\end{equation}
where $\mathbf{\tilde{Z}}$ is the quantized hyper-prior information, $\mu_m$ is the location parameter, and $b_m$ is the scale parameter. Then, $\mathbf{\tilde{Y}}$ and $\mathbf{\tilde{Z}}$ are losslessly encoded using an arithmetic encoder.

At the decoder, the quantized latent representation $\mathbf{\tilde{Y}}$ is mapped to the reconstructed radius matrices $\mathbf{\overline{A}}$. The proposed VRC module is trained with RD loss:
\begin{equation}
\mathcal{L}_{r-d} = R + \lambda \cdot D, 
\end{equation}
where $\lambda$ is a trade-off parameter. The rate $R$ is estimated as
\begin{equation}
R = -\frac1{M}\sum_{m=1}^{M}{\log_2 p(\tilde{y}_m | \mathbf{\tilde{z}})}-\frac1{T}\sum_{t=1}^{T}{\log_2 p(\tilde{z}_t)}, 
\end{equation}
where $M$ is the number of elements in $\mathbf{\tilde{Y}}$, $T$ is the number of elements in $\mathbf{\tilde{Z}}$, and $\mathbf{\tilde{z}}$ is an elements in $\mathbf{\tilde{Z}}$. The distortion $D$ is expressed as
\begin{equation}
D = \frac{1}{256^2} \sum_{i=1}^{256} \sum_{j=1}^{256} \left\| \mathbf{A}_{i,j} - \mathbf{\overline{A}}_{i,j} \right\|_2^2, 
\end{equation}

\subsection{RD-Optimized Quantization Parameter Selection}
Since each coordinate axis in the spherical coordinate system has a different physical meaning, the distortion caused by the quantization of residuals on each axis contributes differently to the overall reconstruction distortion. Here, we use meters ($m$) as the unit for the radius and degrees ($^\circ$) for the angles and analyze the impact of distortions in $\phi,\theta$ and $r$ on the distortion of the reconstructed point cloud. 

We first assume that only the radius $r$ has a reconstruction error of $\Delta r$. The distortion $\Delta d_r$ in Cartesian coordinates caused by $\Delta r$ is expressed as
\begin{equation}
\Delta d_r = \frac{\Delta r}{\cos\theta}. 
\end{equation}
Then, we assume that only the elevation angle $\theta$ has a reconstruction error of $\Delta \theta$. The reconstructed Cartesian coordinates $(\overline{x},\overline{y},\overline{z})$ can be expressed as
\begin{equation}
\begin{aligned}
\overline{x} &= r \cos \phi \\
\overline{y} &= r \sin \phi \\
\overline{z} &= r \tan(\theta + \Delta \theta),
\end{aligned}
\end{equation}
and the distortion $\Delta d_\theta$ caused by $\Delta \theta$ can be expressed as
\begin{equation}
\Delta d_\theta = \sqrt{(x-\overline{x})^2+(y-\overline{y})^2+(z-\overline{z})^2}. 
\end{equation}
By combining (25) and (26), we obtain
\begin{equation}
\Delta d_\theta =  \frac{r\sqrt{2(1-\cos\Delta \theta)}}{\cos(\theta+\Delta \theta)}.
\end{equation}
Similarly, if we assume that only the azimuth angle $\phi$ has a reconstruction error of $\Delta \phi$, the reconstructed Cartesian coordinates $(\overline{x},\overline{y},\overline{z})$ can be expressed as
\begin{equation}
\begin{aligned}
\overline{x} &= r \cos(\phi + \Delta \phi) \\
\overline{y} &= r  \sin(\phi + \Delta \phi) \\
\overline{z} &= r \tan(\theta),
\end{aligned}
\end{equation}
By combining (28) and (26), the distortion $\Delta d_\phi$ due to $\Delta \phi$ can be expressed as
\begin{equation}
\Delta d_\phi = r\sqrt{2(1-\cos\Delta \phi)}.
\end{equation}
As shown in (24), (27), and (29), the distortion on each axis contributes differently to the reconstruction distortion in spherical coordinates. if $\Delta \phi=\Delta \theta$, then $\Delta d_\theta\geq\Delta d_\phi$, indicating that $\Delta \theta$ has a greater impact than $\Delta \phi$. Because LiDAR point cloud geometry information typically requires lossless or near-lossless compression, the mean angular distortion is generally less than $0.1^\circ$. Therefore, for easier comparison between $\Delta d_\theta$ and $\Delta d_r$, we assume $\Delta \theta=0.1^\circ$ and $\Delta r=0.1m$. This results in $\Delta d_r=0.1m$ and $\Delta d_\theta\approx 0.0015r$. Consequently, $\Delta d_\theta\geq\Delta d_r$ only in the rare case when $r\geq 66.7m$.

\begin{table}
\begin{center}
  \caption{Bits consumption of $\phi,\theta,r$, as well as their percentages in the total coding bits}
  \vspace{-0.2cm}
  \label{tab1}
  \begin{tabular}{cccc}
    \toprule
     &$\phi$ &$\theta$ &$r$\\
    \midrule
    Bitrate&1.03& 1.28 & 2.23  \\
    Percentage&22.7\%& 28.2\% &  49.1\%  \\
  \bottomrule
\end{tabular}
\end{center}
\vspace{-0.6cm}
\end{table}

On the other hand, due to the different distributions of $\phi,\theta$ and $r$, the coding efficiencies of $\phi$ and $\theta$ are higher than that of $r$. Specifically, as shown in Fig. 3, the elevation angles of points with the same laser ID typically vary within a range of no more than $2^\circ$ and exhibit a negative correlation with the radius. Therefore, the elevation angle is easy to predict, achieving high encoding efficiency. Although the azimuth angle ranges from $0^\circ$ to $360^\circ$, the differences in azimuth angles between adjacent points in predictive tree are often close to the LiDAR's inherent azimuth resolution (see (7)). Accordingly, the azimuth angle can also be efficiently encoded. However, the radius typically ranges from 0 to 100 $m$ and lacks clear variation patterns, resulting in inefficient compression. For example, Table 1 compares the bitrates of encoding $\phi,\theta$ and $r$, where $q_{\phi}=1$, and $q_\theta$ and $q_r$ are both set to $16$. We can see that the bitrate of the radius accounts for half of the total bitrate and is significantly higher than that of the elevation and azimuth angles.

In summary, $\Delta r$ has a significant influence on distortion for most points but shows low encoding efficiency, while $\Delta \phi$  and $\Delta \theta$ have less impact on distortion but demonstrate high encoding efficiency. Furthermore, the nonlinear relationship between different coordinate systems makes it difficult to calculate the best QPs. Therefore, we propose a DE-based algorithm to select optimal QPs.

In high-bitrate coding mode, there are four QPs: $q_\delta$, $q_{\phi}$, $q_\theta$, and $q_r$, whereas in low-bitrate coding mode, there are three QPs: $q_\delta$, $q_{\phi}$, and $q_\theta$. Here, we use high-bitrate encoding as an example to introduce the proposed QP selection method. Specifically, we first model the QPs selection problem as an optimization problem under a given bitrate constraint:
\begin{equation}
 \begin{gathered}
\min_{q} \quad\sum_{n=1}^N D(\mathbf{q}, \mathbf{P}_n, \mathbf{\overline{P}}_n) \\
\mathbf{s.t.} \quad \frac{1}{N}\sum_{n=1}^N R(\mathbf{q}, \mathbf{P}_n) \leq R_T,
\end{gathered}
\end{equation}
where $\mathbf{q}=(q_\delta, q_{\phi}, q_\theta,q_r)$ is the set of QPs, $D$ is the MSE between reconstructed point cloud $\mathbf{\overline{P}}_n$ and the original point cloud $\mathbf{P}_n$, $R(\mathbf{q}, \mathbf{P}_n)$ is the bitrate for encoding $\mathbf{P}_n$, $N$ is the number of point clouds, and $R_T$ is the target bitrate.

We apply the DE algorithm to a small dataset containing only $20$ point clouds to derive a solution $\mathbf{q}^{*}$ during training. Specifically, we first randomly initialize the candidate solutions, which are also called the population in the DE algorithm
\begin{equation}
 \begin{gathered}
\mathbf{Q}(t) = (\mathbf{q}_{1}(t),\mathbf{q}_{2}(t),...,\mathbf{q}_{K}(t)) \\
\mathbf{q}_{k}(t)=(q_{\delta,k}(t), q_{\phi,k}(t), q_{\theta,k}(t), q_{r,k}(t)),
\end{gathered}
\end{equation}
where $\mathbf{Q}(t)$ represents the population after $t$ iterations, $\mathbf{q}_{k}(t)$ is an individual in $\mathbf{Q}(t)$, and $K$ is the population size. To enhance the efficiency of the DE algorithm, the range of the QPs is restricted to:
\begin{equation}
 \begin{gathered}
q_{\delta}, q_{\theta}, q_{r} \in \left\{x \,\middle|\, x \in \mathbb{Z}, \, 1 \leq x \leq 256 \right\}, \\
q_{\phi} \in \left\{ x \,\middle|\, x \in \mathbb{Z}, \, 1 \leq x \leq 16 \right\}.
\end{gathered}
\end{equation}
The fitness function $f$ of $\mathbf{q}_{k}(t)$ is defined as
\begin{equation}
 \begin{gathered}
f_k(t) = \sum_{n=1}^N D(\mathbf{q}_{k}(t),\mathbf{P}_n, \mathbf{\overline{P}}_n).
\end{gathered}
\end{equation}
The fitness and bitrate of each individual in the population are evaluated. For individuals that do not satisfy the constraint $\frac{1}{N}\sum_{n=1}^N R(\mathbf{q}_{k}(t), \mathbf{P}_n) \leq R_T$, their fitness values are assigned positive infinity. For simplicity, this constraint is abbreviated as $R(\mathbf{q})\leq R_T$.

To maintain the diversity of the population, the DE algorithm introduces perturbations by leveraging the differences between individuals. The perturbation $\mathbf{b}_{k,l}(t)$ is expressed as
\begin{equation}
\mathbf{b}_{k,l}(t) = \mathbf{q}_{k}(t) - \mathbf{q}_{l}(t),
\end{equation}
which is weighted and then added to another individual $\mathbf{q}_{j}(t)$ to obtain the mutant individual $\mathbf{v}_{j}(t)$:

\begin{equation}
\mathbf{v}_{j}(t+1) = \mathbf{q}_{j}(t) + \mu \mathbf{b}_{k,l}(t),
\end{equation}
where $\mu\in [0,2]$ is called the scale factor. The QP in $\mathbf{v}_{j}(t+1)$ is set to the maximum value of the range if it exceeds the preset range in (32), or to the minimum value if it is below the minimum value of the range. To ensure the randomness of mutation, the population size $K$ must not be less than 4; otherwise, mutation cannot be performed\cite{53}.

After mutation, the mutant individual $\mathbf{v}_{j}(t+1)$ and the target individual $\mathbf{q}_{j}(t)$ exchange elements according to the crossover rate $CR$ to form the trial individual $\mathbf{u}_{j}(t+1)$. To ensure the evolution of $\mathbf{q}_{j}(t)$, at least one element in $\mathbf{u}_{j}(t+1)$ must come from $\mathbf{v}_{j}(t+1)$. 

To get better candidate solutions, the individuals in the next generation $\mathbf{q}_{j}(t+1)$ are determined based on the fitness of $\mathbf{q}_{j}(t)$ and $\mathbf{u}_{j}(t+1)$, as well as whether the bitrate constraint is satisfied. The individuals in the next generation can be expressed as
\begin{align}
\mathbf{q}_{j}(t+1) =
\begin{cases} 
\mathbf{u}_{j}(t+1), & f(\mathbf{u}_{j}(t+1)) < f(\mathbf{q}_{j}(t)) \\
& \text{and} \quad R(\mathbf{u}_{j}(t+1))\leq R_T, \\
\mathbf{q}_{j}(t), & \text{otherwise}.
\end{cases}
\end{align}

Through multiple iterations, we obtain $\mathbf{q}^{*}$ under the bitrate constraint and then use it to compress all point clouds. As shown in (24), (27), and (29), since the relationship between distortion on each coordinate axis and reconstruction distortion is invariant, $\mathbf{q}^{*}$ achieved near-optimal RD performance for all point clouds. As $q_\delta$ in $\mathbf{q}^{*}$ is always set to 1 for each bitrate, and the azimuth resolution of LiDAR is generally less than $0.18^\circ$, we ignored the encoding for $\delta$. 

$\mathbf{q}^{*}$ at different bitrates is shown in Table 2, where $r01$ and $r02$ are used for the low-bitrate coding mode and $r03$ to $r07$ are used for the high-bitrate coding mode. In this way, QPs can be obtained through Table 2 without increasing the encoding time.

\begin{table}[t]
    \centering
    \begin{threeparttable}
        \caption{Quantization parameters for each bitrate}
        \label{tab1}
        \begin{tabular}{cccccccc}
            \toprule
             & $r01$ & $r02$ & $r03$ & $r04$ & $r05$ & $r06$ & $r07$ \\
            \midrule
            $q_\phi$    & 1  & 1  & 2  & 2  & 3   & 4   & 8   \\
            $q_\theta$  & 1  & 2  & 2  & 4  & 6   & 12  & 21  \\
            $q_r$       & -   & -   & 12 & 28 & 40  & 81  & 130 \\
            \bottomrule
        \end{tabular}
        \begin{tablenotes}
            \footnotesize
            \item Note: Bitrates for \textit{SemanticKITTI} are ($r01=1.17$, $r02=2.08$, $r03=4.74$, $r04=5.89$, $r05=6.97$, $r06=8.06$, $r07=10.59$) and for \textit{Ford} are $r01=2.34$, $r02=3.46$, $r03=4.71$, $r04=5.45$, $r05=6.44$, $r06=8.22$, $r07=10.51$)
        \end{tablenotes}
    \end{threeparttable}
    \vspace{-0.5cm}
\end{table}

\section{Experiments}
\subsection{Datasets}
We conducted experiments on the LiDAR benchmark \textit{SemanticKITTI} \cite{48} and MPEG-specified dataset \textit{Ford} \cite{49}. \textit{SemanticKITTI} consists of 43552 point clouds and 454.9 million points collected from various environments such as urban areas, countryside, and highways. We applied the default split method using sequences 00 to 10 for training and 11 to 21 for evaluation. The \textit{Ford} dataset is used in MPEG point cloud compression standardization. We used sequence 01 of \textit{Ford} for training and sequences 02 and 03 for evaluation.

\subsection{Implementation Details}
We implemented the proposed LPCM in PyTorch. We trained both the LSTM-P and VRC modules using the Adam optimizer with a learning rate of 0.001 and a learning rate decay $\Gamma=0.99$. We trained the LSTM-P module for 50 epochs with a batch size of 256 and 50 input points. We trained the VRC module for 200 epochs with a batch size of 8. The lambda values $\lambda$ of (20) belongs to \{0.6, 2.2\}.

\begin{figure*}[!t]\centering
  \includegraphics[width=17cm]{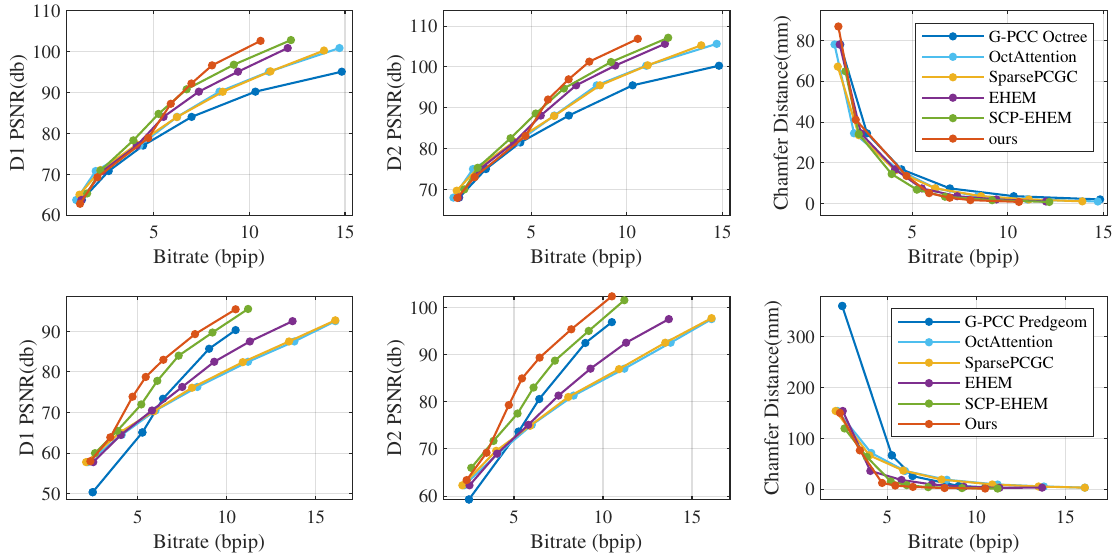}
  \vspace{-0.2cm}
  \caption{RD performance of the methods on the \textit{SemanticKITTI} (top) and \textit{Ford} (bottom) datasets. The horizontal axis denotes bitrate measured in bits per input point. The vertical axis individually represents D1 PSNR (left), D2 PSNR (middle), and CD (right), respectively. In the proposed method, the two points with the lowest bitrates are encoded using the low-bitrate coding mode, while the remaining points are encoded using the high-bitrate coding mode.}
  \label{fig4}
  \vspace{-0.3cm}
\end{figure*}

For the DE algorithm, we set the population size $k$ to 10, the mutation scale factor $\mu$ to 0.4, the crossover rate $CR$ to 0.9, and the number of iterations to 50. As LiDAR point clouds typically require lossless or near-lossless compression to meet the demands of downstream tasks such as autonomous driving and machine perception, we conducted experiments with the smallest $R_T=1.14$. We used $r01$ and $r02$ for low-bitrate coding mode and $r03$ to $r07$ for high-bitrate coding mode.

We conducted all experiments on an Intel(R) Xeon(R) Gold 6148 CPU and a GeForce RTX 4090 GPU with 24 GB of memory. Since the point cloud is divided into multiple predictive trees with no dependencies between them, we used parallel computation to accelerate the process.

\subsection{Baselines}
We verified the effectiveness of the proposed method by comparing it with the state-of-the-art learning-based LiDAR point cloud compression method SCP-EHEM \cite{25} and the G-PCC reference software TMC13 v23.0 \cite{50}. We also compared the proposed method with the representative methods EHEM \cite{23}, SparsePCGC \cite{36} and OctAttention \cite{21} for a more extensive evaluation. As G-PCC does not provide configuration files with calibration parameters for compressing the \textit{SemanticKITTI} dataset, we first represented the point clouds in \textit{SemanticKITTI} using octrees with depths ranging from 10 to 15. We then losslessly compressed the voxelized point clouds using an octree-based encoding method. In contrast, we compressed the point clouds in the \textit{Ford} dataset using the G-PCC predictive geometry encoding method. 
\subsection{Evaluation Metrics}
We used the point-to-point Peak Signal-to-Noise Ratio (D1 PSNR), point-to-plane Peak Signal-to-Noise Ratio (D2 PSNR), and Chamfer distance (CD) to measure the distortion. We measured the bitrate using bits per input point (bpip). Then, we used BD-Bitrate \cite{51} to quantatively compare the RD performance of the proposed method with other methods.

We also implemented and evaluated the proposed method on a downstream task, specifically vehicle detection. For this task, we used Voxel-R-CNN \cite{52} as the detector and calculated the average precision (AP) to assess detection accuracy on the reconstructed point clouds. We set the mean intersection-over-union (mIOU) threshold to 0.7, which is a standard benchmark for evaluating object detection algorithms.

\begin{table*}
\centering
\caption{BD-Rate gains on the \textit{Ford} dataset}
\label{tab:bd_rate}
\begin{tabular}{lccccccccc}
\toprule
\multirow{2}{*}{Method} & \multicolumn{3}{c}{High-bitrate coding mode} & \multicolumn{3}{c}{Low-bitrate coding mode} & \multicolumn{3}{c}{All bitrates} \\  
 &   D1 PSNR & D2 PSNR&CD&D1 PSNR &D2 PSNR & CD& D1 PSNR &D2 PSNR & CD \\ \midrule
G-PCC &  -23.5\% & -21.9\% & -29.0\% & -28.6\% & -21.2\% & -34.7\% &  -24.9\% & -21.7\% & -30.6\%\\
SparsePCGC &  -41.1\% & -44.2\% & -40.3\% & 3.4\% & -4.8\% & 5.1\%&  -28.4\% & -33.2\% & -27.7\% \\
OctAttention &  -42.2\% & -45.3\% & -41.4\% & -4.7\% & -13.4\% & -6.1\% &  -31.5\% & -36.2\% & -31.3\%\\
EHEM &  -32.5\% & -35.8\% & -32.7\% & -8.2\% & -15.3\% & -9.3\%&  -25.6\% & -29.9\% & -26.0\% \\
SCP-EHEM &  -10.1\% & -12.9\% & -11.7\% & 1.3\% & 2.4\% & 1.2\% &  -6.8\% & -8.5\% & -8.0\%\\ \bottomrule
\end{tabular}
\vspace{-0.3cm}
\end{table*}

\begin{table*}
\centering
\caption{BD-Rate gains on the \textit{SemanticKITTI} dataset}
\label{tab:bd_rate}
\begin{tabular}{lccccccccc}
\toprule
\multirow{2}{*}{Method} & \multicolumn{3}{c}{High-bitrate coding mode} & \multicolumn{3}{c}{Low-bitrate coding mode} & \multicolumn{3}{c}{All bitrates} \\  
 &   D1 PSNR & D2 PSNR&CD&D1 PSNR &D2 PSNR & CD& D1 PSNR &D2 PSNR & CD \\ \midrule
G-PCC &  -29.9\% & -30.5\% & -41.5\% & -6.2\% & -7.7\% & -6.8\% &  -23.1\% & -23.9\% & -31.6\%\\
SparsePCGC &  -21.3\% & -20.4\% & -15.8\% & 10.4\% & 11.6\% & 8.2\%&  -12.2\% & -11.2\% & -8.9\% \\
OctAttention &  -22.3\% & -21.5\% & -13.0\% & 8.8\% &6.8\% & 8.3\% &  -13.3\% & -13.4\% & -6.9\%\\
EHEM &  -10.7\% & -9.9\% & -4.8\% & 0.4\% & -1.5\% & -0.1\%&  -7.5\% & -6.6\% & -2.8\% \\
SCP-EHEM &  -1.2\% & -3.0\% & -1.6\% & -3.5\% & -1.3\% & -2.7\% &  -1.9\% & -2.5\% & -1.9\%\\ \bottomrule
\end{tabular}
\vspace{-0.3cm}
\end{table*}

\subsection{Experiment Results}
As shown in Fig. 6, LPCM significantly outperformed all other methods when the bitrate exceeded 5 bpip, while maintaining performance comparable to the state-of-the-art method SCP-EHEM at lower bitrates. Table 3 and Table 4 show the RD performance on the \textit{Ford} and \textit{SemanticKITTI} datasets. We compared the high-bitrate coding mode with the highest five bitrates of the other methods and the low-bitrate coding mode with the lowest two bitrates of the other methods. On the \textit{Ford} dataset, LPCM achieved a D1-based BD-Bitrate reduction of 23.5\% in high-bitrate coding mode, 28.6\% in low-bitrate coding mode, and 24.9\% overall compared to the G-PCC predictive geometry encoding method. Compared with the state-of-the-art method SCP-EHEM, LPCM achieved a D1-based BD-Bitrate in -10.1\% in high-bitrate coding mode and -6.8\% overall.  

On the \textit{SemanticKITTI} dataset, LPCM achieved a D1-based BD-Bitrate of - 29.9\% in high-bitrate coding mode, -6.2\% in low-bitrate coding mode, and -23.1\% overall compared to the G-PCC octree encoding method. Compared to the state-of-the-art SCP-EHEM, LPCM achieved a D1-based BD-Bitrate of -1.2\% in high-bitrate coding mode, -3.5\% in low-bitrate coding mode, and -2.5\% overall across the total bitrate range. The tables also show that the proposed method significantly outperformed the other methods in terms of D2-based BD-Bitrate and CD metrics.

On the \textit{Ford} dataset, our method outperformed the other methods by a greater margin than on the \textit{SemanticKITTI} dataset. This enhanced performance is due to the Ford dataset's higher acquisition precision, enabling our method to more effectively use LiDAR's inherent angular resolution. However, the performance gain was smaller at lower bitrates. This is due to increased reconstruction distortion, which weakens coordinate correlations and hinders accurate geometry prediction.

\begin{figure*}[!t]\centering
  \includegraphics[width=16cm]{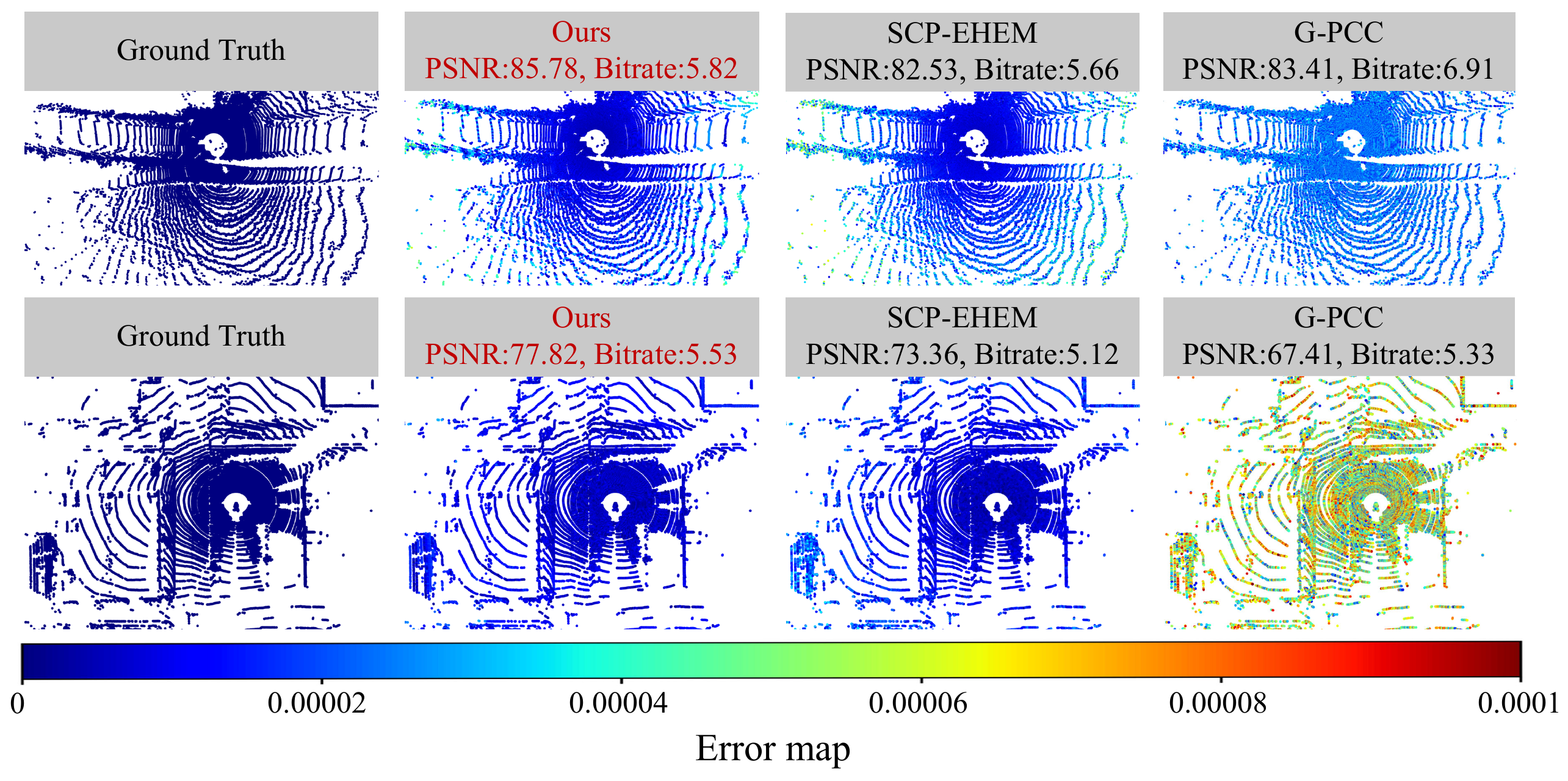}
  \vspace{-0.3cm}
  \caption{Visualization of point clouds reconstructed from G-PCC, SCP-EHEM, and the proposed method on the \textit{SemanticKITTI} (top) and \textit{Ford} (bottom) datasets at similar bitrate. Each result is color-mapped to represent the error, with the color bar indicating the error magnitude.}
  \label{fig4}
  \vspace{-0.4cm}
\end{figure*}

We also present visualizations of the reconstructed point clouds for G-PCC, SCP-EHEM, and the proposed method in Fig. 7. The darker blue regions in the color map indicate smaller distortion, while the darker red regions correspond to larger distortion. The results show that the proposed method achieved higher PSNR than both G-PCC and SCP-EHEM at similar bitrates. The distortion distribution of the point clouds decoded by G-PCC was uniform, whereas the distortion of points with smaller radius in the proposed method was generally smaller than that of points with larger. This is because G-PCC uses octree or radius-based angle-adaptive quantization methods, while our approach derives a globally unified set of quantization parameters through DE. In application scenarios such as autonomous driving, nearby objects play a more significant role. Therefore, we believe that reducing distortion in the near field is reasonable.

\begin{table}
\begin{center}
  \caption{Runtime and model size comparison at high bitrate}
  \label{tab1}
  \begin{tabular}{cccc}
    \toprule
     &Enc. &Dec. &Model size\\
     &(s/frame) &(s/frame) &(MB)\\
    \midrule
    G-PCC & 1.03 & 0.66 &- \\
   OctAttention & 0.71 & 689.24 &28.0\\
    SCP-EHEM  &3.77 &4.02  &329  \\
    High-bitrate coding mode  &3.35 &3.86& 1.35  \\
  \bottomrule
\end{tabular}
\end{center}
\vspace{-0.5cm}
\end{table}

\begin{table}
\begin{center}
  \caption{Runtime and model size comparison at low bitrate}
  \label{tab1}
  \begin{tabular}{cccc}
    \toprule
     &Enc. &Dec. &Model size\\
     &(s/frame) &(s/frame) &(MB)\\
    \midrule
    G-PCC & 0.25 & 0.16 &- \\
   OctAttention & 0.08 & 79.2  &28.0\\
    SCP-EHEM  &0.59 &0.61  &329  \\
    Low-bitrate coding mode  &1.23 &1.56& 4.45  \\
  \bottomrule
\end{tabular}
\end{center}
\vspace{-0.7cm}
\end{table}

The runtime and model size comparisons at similar bitrates are shown in Table 5 and Table 6. In high-bitrate coding mode, we reduced the coding time by using a lightweight model and coding multiple predictive trees in parallel. However, the reconstructed radius $\overline{r}_n$ and reconstructed azimuth angle $\overline{\phi}_n$ require point-by-point prediction. As a result, the coding time of the proposed method was similar to that of SCP-EHEM. In low-bitrate coding mode, since $\phi,\theta$, and $r$ were encoded separately, and the precision of the residuals was reduced, the coding time was shorter compared to the high-bitrate coding mode.

\begin{figure}[!t]\centering
  \includegraphics[width=8cm]{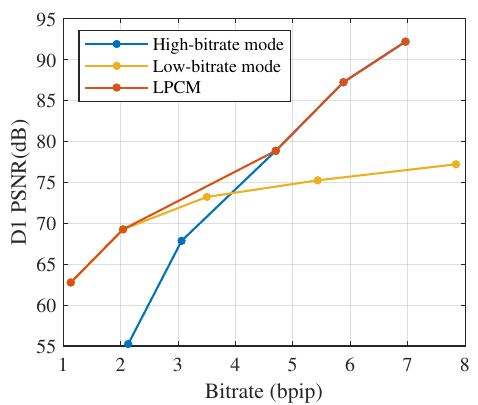}
  \vspace{-0.3cm}
  \caption{Rate-D1 PSNR curves of the high-bitrate coding mode and low-bitrate coding mode.}
  \label{fig8}
  \vspace{-0.6cm}
\end{figure}

\subsection{Ablation Study and Analysis}
We assessed the effectiveness of the high-bitrate coding mode, low-bitrate coding mode, and the impact of the number of input points to the high-bitrate coding mode on the RD performance.

RD performances of the complete proposed method, high-bitrate coding mode at low bitrates, and low-bitrate coding mode at high bitrates are shown in Fig. 8. At low bitrates, the low-bitrate coding mode outperformed the high-bitrate coding mode. This is because the significant distortion at low bitrates made the high-bitrate coding mode unable to accurately predict elevation angles and radii. At high bitrates, the high-bitrate coding mode outperformed the low-bitrate coding mode. As distortion decreased, the LSTM-P module was able to capture the correlation between different coordinates. In addition, the low-bitrate coding mode encoded the radius using an autoencoder, making it challenging to achieve near lossless compression.

We used the BD-Bitrate to compare the RD performance when varying the number of input points provided to the LSTM-P module, with 50 input points as the baseline. As the number of input points exceeded 15, the performance of the LSTM-P module improved. This is because more context information helps the LSTM-P module network better capture the correlation between different coordinates. However, when the number of input points reached 75 and continued to increase, the performance of LSTM-P module began to decline. This suggests that including data from more distant points does not enhance the prediction accuracy for the current point; in fact, it has the opposite effect, leading to a decline in accuracy.

\subsection{Downstream Applications}
Table 8 compares the performance on the vehicle detection task. We evaluated the detection accuracy of point clouds reconstructed using the proposed method, G-PCC, and SCP-EHEM across three detection difficulty levels: easy, moderate, and hard, at different bitrates. The proposed method achieved comparable detection accuracy to the other methods at low bitrates but significantly outperformed them at high bitrates.

From high to low bitrates, the detection accuracy of point clouds reconstructed by G-PCC decreased only slightly. In contrast, the detection accuracy of both our method and SCP-EHEM declined significantly as the bitrate decreases. This is because both the proposed method and SCP-EHEM introduce greater distortion especially in distant regions, which affects detection accuracy. However, even at the lowest bitrate, the proposed method still achieves detection accuracy similar to G-PCC.

\begin{table}
\begin{center}
  \caption{BD-Bitrates of the LSTM-P module. The baseline is LSTM-P with 50 points}
  \label{tab1}
  \begin{tabular}{cccccccc}
    \toprule
    Number of points &15 &30 &75&100\\
    \midrule
    \textit{SemanticKITTI} & 0.33\% & 0.12\% &0.82\%&1.9\%\\
    \textit{Ford} & 0.63\% & 0.34\% &0.62\% &1.5\%\\
    Average & 0.48\%&0.23\%  &0.72\% &1.7\%\\
  \bottomrule
\end{tabular}
\end{center}
\vspace{-0.5cm}
\end{table}

\begin{table}
\centering
\caption{Accuracy  of vehicle detection at different bitrates.}
\begin{tabular}{lcccc}
\toprule
\multirow{2}{*}{Method} & \multirow{2}{*}{Bitrate} &\multicolumn{3}{c}{Detection AP} \\
& & Easy & Mod &Hard\\
\midrule
Raw Data & - & 93.27 & 85.59 & 80.74 \\
\midrule
\multirow{3}{*}{G-PCC} & 2.63 & 85.71 & 78.73 & 75.98 \\

 & 4.43 & 86.94 & 78.31 & 76.17 \\

 & 6.97 & 90.04 & 79.83 & 77.09 \\
\midrule
\multirow{3}{*}{SCP-EHEM} & 2.27 & 85.40 & 79.03 & 76.33 \\

 & 4.12 & 88.28 & 79.13 & 76.67 \\

 & 6.71 & 92.18 & 84.08 & 78.93 \\
\midrule
\multirow{3}{*}{Ours} & 2.04 & 85.54 & 78.67 & 76.23 \\

 & 4.70 & 89.99 & 80.19 & 77.35 \\

 & 6.96 & \textbf{93.20} & \textbf{85.72} & \textbf{80.18} \\
\bottomrule
\end{tabular}
\vspace{-0.5cm}
\end{table}

\section{Conclusions}
We introduce LPCM, a learning-based predictive coding framework specifically designed for LiDAR point cloud compression. Our method aims to overcome the limitations of existing approaches, which fail to exploit long-term geometric correlations and do not optimize quantization parameters effectively within the spherical coordinate. The high-bitrate coding mode uses the LSTM-P module to model correlations across different coordinates, enabling accurate prediction of elevation angles. In contrast, the low-bitrate coding mode uses the VRC module to efficiently compress the radii of points directly. Moreover, we introduced a DE-based QP selection method within the spherical coordinate system, which enhanced RD performance without incurring additional coding time. Experimental results showed that LPCM outperformed existing methods in RD performance and excelled in a vehicle detection task.

\vspace{-1cm}

\vfill

\end{document}